\documentclass[twocolumn,showpacs,pre,letterpaper,final]{revtex4}
\usepackage{amscd}
\usepackage{amssymb}
\usepackage{amsfonts}
\usepackage{amsmath}
\usepackage{graphicx}
\usepackage{dcolumn}
\usepackage[dvipdfm]{hyperref}

\begin{document}

\preprint{}
\title{Rich-club connectivity dominates assortativity and transitivity of complex networks}
\author{Xiao-Ke Xu $^{1,2}$}
\email{xuxk@eie.polyu.edu.hk}
\author{Jie Zhang $^{3,2}$}
\author{Michael Small $^2$}
\affiliation{$^1$School of Communication and Electronic
Engineering, Qingdao Technological University, Qingdao 266520, People's Republic of China\\
$^2$Department of Electronic and Information Engineering, Hong Kong
Polytechnic University, Hong Kong, People's Republic of China \\
$^3$Centre for Computational Systems Biology, Fudan University,
Shanghai 200433, People's Republic of China}
\date{\today}

\begin{abstract}
Rich-club, assortativity and clustering coefficients are
frequently-used measures to estimate topological properties of
complex networks. Here we find that the connectivity among a very
small portion of the richest nodes can dominate the assortativity
and clustering coefficients of a large network, which reveals that
the rich-club connectivity is leveraged throughout the network. Our
study suggests that more attention should be payed to the
organization pattern of rich nodes, for the structure of a complex
system as a whole is determined by the associations between the most
influential individuals. Moreover, by manipulating the connectivity
pattern in a very small rich-club, it is sufficient to produce a
network with desired assortativity or transitivity. Conversely, our
findings offer a simple explanation for the observed assortativity
and transitivity in many real world networks --- such biases can be
explained by the connectivities among the richest nodes.
\end{abstract}


\maketitle
After ten years of explosive growth, fruitful measures based on
statistical physics have been proposed for analyzing all kinds of
complex networks \cite{Review_measure}. Measures such as degree
distribution, average degree, clustering coefficient, assortativity
coefficient, and average shortest-path length, are now widely used
in almost all complex networks to estimate their topological
properties. For example, clustering coefficient \cite{SW} is used to
measure the transitivity property of a network. If a social network
has a high clustering coefficient, it means that the friends of
someone are also likely to be friends themselves
\cite{Social_mixing}.

A second popular measure is the assortativity coefficient which
defines the mixing pattern among the nodes. A positive coefficient
indicates that nodes with similar degrees tend to be connected to
each other (assortative mixing), while a negative coefficient
captures the opposite case in which very different degree nodes are
connected (disassortative mixing) \cite{Social_mixing,
Mixing_Newman02}. Although the above calculations on assortativity
and transitivity may be useful in many situations, the actual
validity of these measures to capture the true assortativity and
transitivity of the network has not been verified. In particular,
the effectiveness of assortativity coefficient in some specific
networks has been critically examined  recently \cite{AS_Xu,
AS_PRL}.

Many real networks display a skewed degree distribution \cite{BA},
so a small number of nodes possess much higher degrees than the
overwhelming majority. Nonetheless, it is necessary to be cautious
in applying such statistical measures as the actual value of most
statistics (e.g., assortativity and clustering coefficients) is the
statistical average of a whole network, and this averaging process
may conceal the prominent effect of the richest elements
\cite{Weight_richclub}. Furthermore, it is already clear that the
small number of rich nodes play a central role in static and dynamic
processes on complex networks, such as targeted attack
\cite{Attack}, cascade failure \cite{Cascading_failure}, and disease
spreading \cite{Super_spreader}. Therefore, more attention should be
paid to rich nodes when analyzing finite-size network data
\cite{AS_Xu}. In particular, it is interesting to analyze the
organization pattern of rich nodes \cite{Richnode_PA}, such as
whether rich nodes trend to connect to one another, or with the rest
of nodes \cite{Richclub_origin}.

\begin{table*}[htbp]
\centering \caption{Statistics of nine undirected networks: number
of nodes $n$, average degree $\langle k\rangle$, the exponent of
degree distribution if the distribution follows a power law:
$\alpha$ (or ``--'' if not) , structural cutoff degree
$k_s=\sqrt{\langle k\rangle n}$ \cite{Degree_cutoff}, maximal degree
$k_{max}$, assortativity coefficient $r$ \cite{Mixing_Newman02},
clustering coefficient $c$ \cite{SW}, and average shortest-path
length $l$. SW is the network generated by the small-world model
\cite{SW}, ER is the network generated by Erd\H{o}s-R\'{e}nyi model
\cite{ER}, PG is the network of US power grid \cite{BA}, COND is the
network of scientists who work on condensed matter
\cite{Newman_datasets}, BA is the network generated by the
scale-free model \cite{BA}, EPA is the network from the pages
linking to www.epa.gov \cite{Pajek_data}, PFP is the network
generated by the model for the Internet topology \cite{PFP}, AS is
the network of the Internet topology at the level of autonomous
systems \cite{ASdata} and BOOK is the word adjacency network of text
from Darwin's ``The Origin of Species'' \cite{Superfamily}. The
proportion of rich nodes in all the networks is $0.5\%$ except the
network of COND. We select less proportion ($0.2\%$) nodes as rich
nodes in COND, because it has larger scale (more nodes) than other
networks. For $r$, $c$ and $l$, the first row is the value when rich
nodes do not connect to other rich nodes (without rich-club), and
the second row is the value when rich nodes completely connect to
each other (with rich-club). }
\begin{ruledtabular}
\begin{tabular}{ c c c c c c c c c c}
Network & SW & ER & PG & COND & BA & EPA & PFP & AS & BOOK\\
\hline
$n$ & $5000$ & $5000$ & $4941$ & $16726$ & $5000$ & $4772$ & $5000$ & $5375$ & $7724$ \\
$\langle k\rangle$ & $6.0$ & $10.0$ & $2.7$ & $5.7$ & $6.0$ & $3.7$ & $6.0$ & $3.9$ & $11.4$ \\
$\alpha$ & $-$ & $-$ & $-$ & $-$ & $3.0$ & $2.0$ & $2.2$ & $2.2$ & $1.9$ \\
$k_{max}$ & $15.5\pm{3.5}$ & $23.4\pm{1.6}$ & $19$ & $107$ & $218.6\pm{43.4}$ & $175$ & $1258.8\pm{349.0}$ & $1193$ & $2568$ \\
$k_s$ & $173.2$ & $223.6$ & $115.4$ & $308.8$ & $173.2$ & $132.9$ & $173.2$ & $144.8$ & $296.7$ \\
{$k_{max}/k_s$} & $0.09$ & $0.10$ & $0.16$ & $0.35$ & $1.26$ & $1.32$ & $7.26$ & $8.24$ & $8.66$ \\
\hline
 & $\textbf{0.00}\pm{\textbf{0.00}}$ & $\textbf{0.00}\pm{\textbf{0.00}}$ & $\textbf{-0.01}$ & $\textbf{0.17}$ & $\textbf{-0.08}\pm{\textbf{0.01}}$ & $\textbf{-0.31}$ & $-0.25\pm{0.04}$ & $-0.19$ & $-0.24$
 \\
\raisebox{1.3ex}[0pt]{$r$} & $\textbf{0.69}\pm{\textbf{0.00}}$ & $\textbf{0.39}\pm{\textbf{0.00}}$ & $\textbf{0.60}$ & $\textbf{0.32}$ & $\textbf{0.04}\pm{\textbf{0.02}}$ & $\textbf{-0.15}$ & $-0.24\pm{0.04}$ & $-0.19$ & $-0.24$ \\
\hline
 & $0.44\pm{0.00}$ & $0.00\pm{0.00}$ & $0.08$ & $0.62$ & $0.00\pm{0.00}$ & $0.04$ & $\textbf{0.15}\pm{\textbf{0.01}}$ & $\textbf{0.10}$ & $\textbf{0.21}$ \\
 \raisebox{1.3ex}[0pt]{$c$} & $0.44\pm{0.00}$ & $0.00\pm{0.00}$ & $0.08$ & $0.62$ & $0.02\pm{0.00}$ & $0.07$ & $\textbf{0.28}\pm{\textbf{0.02}}$ & $\textbf{0.26}$ & $\textbf{0.41}$ \\
\hline
 & $7.85\pm{0.05}$ & $3.94\pm{0.01}$ & $6.63$ & $6.64$ & $4.11\pm{0.02}$ & $4.63$ & $3.17\pm{0.06}$ & $3.95$ & $2.87$ \\
 \raisebox{1.3ex}[0pt]{$l$} & $7.33\pm{0.03}$ & $3.94\pm{0.01}$ & $6.37$ & $6.37$ & $3.94\pm{0.02}$ & $3.97$ & $3.04\pm{0.05}$ & $3.60$ & $2.77$ \\
\end{tabular}
\end{ruledtabular}
\label{table1}
\end{table*}

Compared with a corresponding randomized network, if rich nodes are
interconnected to one another more intensely than to low-degree
nodes, the network is said to have a rich-club property
\cite{Colizza_richclub, NP_comment, Zhou_richclub, APL_richclub,
Constraint}. Note that, rich-club only describes the property of
rich nodes, and it is not a statistical average over the entire
network. Rich-club is therefore different from the statistics that
are based on the averaged results over all nodes (like clustering
and assortativity coefficients). In this study, we demonstrate that
the connections among a very small portion (no more than $0.5\%$) of
rich nodes control the statistical properties of the entire complex
networks, especially assortativity and transitivity properties. We
find that adding a small number of extra links among rich nodes can
significantly increase an assortativity coefficient to be positive,
and raise a low clustering coefficient to a high value. These
results show that it is possible to engineer the transitive or
assortative features of a large complex network just by altering the
wiring structure within a very small rich-club. Finally, this work
allows us to explain the observed assortativity/transitivity of
various real world networks (e.g. the Internet) by studying the
connectivity between the richest nodes. That is, the structure of a
complex system is mostly determined by the associations between the
most influential individuals.

We select the top $0.5\%$ of the highest degree nodes as rich nodes
in a network and manipulate the connections among them. First we
make rich nodes fully connected to one another, so they form a
completely connected rich-club. Secondly, we completely eradicate the
edges among these rich nodes, so that the network has no rich-club.
The topological structure is the same for the above two networks
except for the connection pattern among rich nodes. Then we
calculate the frequently-used statistics for the above two networks
respectively to compare how the absence and presence of a rich-club
affects the statistical properties of the whole network.

Table \ref{table1} lists the results of nine undirected networks
(including five real networks and four model networks) arranged with
$k_{max}/k_s$ increasing. The value of the structural cutoff degree
$k_s$ can be regarded as the first approximation in a scale-free
network \cite{Degree_cutoff}. Here $k_{max}/k_s$ is a convenient
index that can be used in complex networks with any degree
distribution to show the proportion of links (or degrees) the rich
nodes possess in comparison with the rest nodes in a network. Lower
$k_{max}/k_s$ means that the degrees of rich nodes are close to the
majority of nodes, while a high $k_{max}/k_s$ indicates that the
degrees of rich nodes are far larger than the rest.

The results in Table \ref{table1} show whether a very small
proportion of rich nodes forms a club can partly control the two
important statistics: assortativity coefficient $r$ and clustering
coefficient $c$. Based on the different values of $k_{max}/k_s$,
complex networks fall into two distinct groups. In the networks with
low $k_{max}/k_s$ like SW, ER, PG, COND, BA and PG, the values of
$r$ are largely determined by the rich-club. But for the networks
with high $k_{max}/k_s$ such as PFP, AS and BOOK, the values of $c$
are largely determined by the rich-club.

Now we analyze how the rich-club connectivity dominates $r$.
Recently, the effectiveness of $r$ in some specific networks has
been queried. In our previous work \cite{AS_Xu}, we found that
superrich nodes (degree much larger than the natural cutoff value
\cite{Degree_cutoff}) can strongly influence $r$. Meanwhile, another
work showed that the highly heterogeneous (scale-free) network with
``natural'' degree mixing has a disassortative coefficient
\cite{AS_PRL}. These studies indicate that $r$ is always strongly
negative for some specific networks \cite{Constraint}. In Table
\ref{table1}, we also find that $r$ is strongly negative for the
networks with a high $k_{max}/k_s$ (i.e., with superrich nodes
\cite{AS_Xu}), such as PFP, AS and BOOK.

While the above studies focus on the effect of rich nodes, in this
work we pay more attention to how the {\em organization} of rich
nodes (to form a rich-club or not) affects $r$. For networks with
low $k_{max}/k_s$ and the absence of a rich-club such as SW, ER and
PG, the values of $r$ are near zero, which indicates that these
networks are neutral mixing. But the counterparts with the presence
of a rich-club show a surprisingly positive $r$, which implies that
these networks have assortative mixing properties. It is obvious
that the mixing patterns of more than $99.5\%$ nodes remain
unchanged, so this metamorphosis is induced by the absence and
presence of the rich-club. For the networks COND, BA and EPA, our
results again imply that the connections among no more than $0.5\%$
rich nodes can make $r$ become much more positive.

For networks with a high $k_{max}/k_s$, such as PFP, AS and BOOK,
the presence of a rich-club does slightly affect $r$, while it
strongly affects $c$. Traditionally, high $c$ indicates that the
friends of someone are also likely to be friends themselves. A
highly assortative network often implies a high $c$ as nodes with
similar degrees will connect to each other \cite{High_assort} and
form multiscale communities \cite{Social_mixing}. But in a highly
disassortative network, a high-degree node trends to connect to a
low-degree node, which in turn connects to another high-degree node,
and this high-low-high-low connection circle will lead to a low $c$.
It is therefore not obvious why a high $c$ emerges in
disassortative networks like PFP, AS and BOOK.

\begin{figure}[htbp]
\centering
\includegraphics[width=0.5\textwidth]{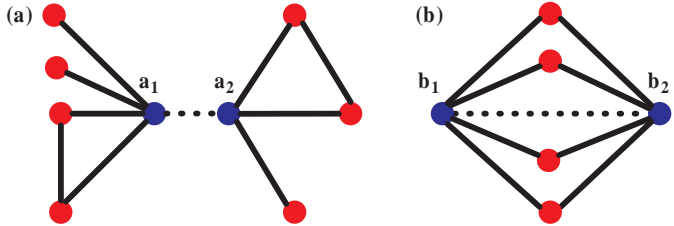}
\caption{(Color online) (a) Whether rich nodes $a_1$ and $a_2$ are
to be connected will not significantly affect clustering coefficient
$c$, while (b) whether rich nodes $b_1$ and $b_2$ form a rich-club
strongly affects $c$.} \label{Clustering}
\end{figure}

Although the high values of $c$ in the high disassortative networks
with rich-club are contrary to our intuition, this phenomenon can be
partly explained by considering the effect of the  rich-club in more
detail. As has been shown in Fig. \ref{Clustering}(a), if rich nodes
$a_1$ and $a_2$ are connected to each other, the value of $c$ for
this network will only change slightly. While if rich nodes $b_1$
and $b_2$ are connected to each other as is shown in Fig.
\ref{Clustering}(b), the network will show a high $c$. Moreover, the
scenario in Fig. \ref{Clustering}(b) shows that a high $c$ does not
always imply that the friends of someone are also likely to be
connected for some specific networks. For example, even if $b_1$
connecting to $b_2$ makes the network in Fig. \ref{Clustering}(b)
show a high $c$, the other four low-degree nodes do not connect to
each other either.

For other statistics such as average degree, degree distribution,
and average shortest-path length, it is easy to guess how the
presence or absence of a rich-club can influence them. Because the
proportion of rich nodes manipulated here is no more than $0.5\%$,
the degree distribution and average degree remain largely unchanged
whether a network has a rich-club or not. Another statistic that is
vulnerable to rich-club phenomena is average shortest-path length
$l$ \cite{Richclub_origin}. Rich nodes often act as a traffic hub
and provide a large selection of shortcuts, hence we can guess that
a network without rich-club may lose the efficiency compared with
its rich-club counterpart. For all the nine networks in Table
\ref{table1}, this conjecture is right, for the presence and absence
of a rich-club also strongly affects $l$, although not as strong as
$r$ and $c$.

It should be noted that a large $k_{max}/k_s$ can reduce $l$ more
significantly than the presence of a rich-club. For networks with
the same average degree, such as SW and PFP in Table \ref{table1},
the degree of the richest node in SW is far lower than that in PFP,
so the value of $l$ in the former is larger than the latter. In the
network with low $k_{max}/k_s$ (SW), every rich node only connects
to a small number of nodes and they can only provide sparse
shortcuts for other nodes, so the network has a longer $l$
$[7.33\sim7.85]$. In the network with high $k_{max}/k_s$ (PFP), rich
nodes have to connect to a huge number of low-degree nodes, so rich
nodes provide a lot of shortcuts to low-degree nodes and the network
has a shorter $l$ $[3.04\sim3.17]$.

\begin{figure}[htbp]
\centering
\includegraphics[width=0.3\textwidth]{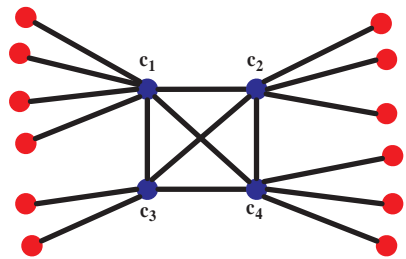}
\caption{(Color online) A toy model to show the dilemma of rich-club
definition \cite{Zhou_richclub}. Rich nodes $c_1$--$c_4$ have larger
degrees and form a subnetwork in which rich nodes are completely
connected to one another, so the network has a rich-club according
to the definition in \cite{Richclub_origin, Constraint}. But there
is no rich-club using the definition in \cite{Colizza_richclub}, for
$c_1$--$c_4$ are always connected to each other too in its
corresponding randomized network.} \label{dilemma}
\end{figure}

Whether a network should be considered as having a rich-club has
been discussed directly in some specific networks. For example,
whether the network of Internet has a rich-club has been debated
\cite{Richclub_origin, Colizza_richclub, Constraint}, and there is
still not a clear conclusion. Furthermore, a dilemma of rich-club
definition occurred in \cite{Zhou_richclub} and is shown in Fig.
\ref{dilemma}. In the definition of Zhou and Mondrag\'{o}n
\cite{Richclub_origin}, they only study whether rich nodes are more
likely to interconnect than to low-degree nodes, so that our toy
model is therefore regarded as having a rich-club. However, Colizza
{\it et al.} believe that rich-club should be inferred by a
comparison of the original network with its randomized counterparts
(reference network) \cite{Correlation_Science} to avoid the false
inference of rich-club in non-rich-club networks. Consequently, for
the toy model in Fig. \ref{dilemma}, the method in
\cite{Colizza_richclub} will run into a dilemma, for the original
network and its randomized version show the same structure.

To harmonize this contradiction, the frequently-used statistics can
be used to judge whether a network has a rich-club. For the network
with low $k_{max}/k_s$, we prefer to use $c$ as the primary
statistic; while for the network with high $k_{max}/k_s$, we can use
$r$ instead. Our framework is based on whether the statistics of the
original network are strongly affected by the absence and presence
of a rich-club. If the statistics of the original network are more
similar to its fully-connected rich-club counterparts, and are far
away to its non-rich-club counterparts, we can conclude that the
network has a rich-club. Conversely, if this is not the case then we
would conclude that the network has no rich-club.

We now use this new method to judge whether the Internet has a
rich-club. We list the statistics $r$, $c$, and $l$ for the four
versions of the Internet network in Table \ref{table2}: the network
without rich-club, the original network, the network with rich-club
and the corresponding randomized network. The properties of the
original network are found to be more close to the network with
rich-club, and are substantially different to the network without
rich-club. This is especially obvious for the value of $c$, so it is
easy to conclude that the network has a rich-club.

\begin{table}[htbp]
\caption{Statistics on four versions of the Internet network at the
level of autonomous systems \cite{ASdata}: the number of total links
among rich nodes $m$, clustering coefficient $c$ \cite{SW},
assortativity coefficient $r$ \cite{Mixing_Newman02}, and average
shortest-path length $l$. We choose $27$ nodes ($0.5\%$ of the whole
nodes) with the highest degrees as rich nodes. Origin stands for the
original network; non-rich-club stands for the original network
deleted the links among rich nodes; rich-club stands for the
original network in which rich nodes are completely connected to
each other; random stands for the randomized version of the original
network generated by the random mixing method
\cite{Correlation_Science}.}
\begin{ruledtabular}
\begin{tabular}{ c c c c c}
Network & non-rich-club & origin & rich-club & random \\
\hline
$m$ & $0$ & $148$ & $351$ & $209.4\pm{10.4}$ \\
$c$ & $0.10$ & $0.24$ & $0.26$ & $0.13\pm{0.00}$ \\
$r$ & $-0.19$ & $-0.18$ & $-0.19$ & $-0.18\pm{0.00}$ \\
$l$ & $3.95$ & $3.70$ & $3.60$ & $3.54\pm{0.01}$ \\
\end{tabular}
\end{ruledtabular}
\label{table2}
\end{table}

Our new method for measuring rich-club can provide a more
satisfactory and impartial judgement on whether a network has a
rich-club. The new method does not depend explicitly on how many
links there are among rich nodes as previous measures that have been
taken \cite{Colizza_richclub}. Rather, our approach is to directly
measure the effect that the rich-club has on the properties of the
whole network. Nonetheless, we are not suggesting that the existing
tools for detecting rich-clubs should be abandoned. The controversy
over whether particular networks have a rich-club is due to the
tension between what are meant with evocative names and description
(as are associated with the term ``rich-club'') and what is actually
being measured with various statistics. A more appropriate question
is what effect these measured properties have on the network
structure and dynamics.

In this work, we focus on how the rich-club affects the basic
statistics of complex networks, especially assortativity and
clustering coefficients. Our findings uncover the effect of the
organization of rich nodes, which leads to a better understanding of
the behavior of a complex system. These results show that just by
altering the wiring structure within a very small rich-club one can
engineer the transitive or assortative features of a large complex
network. The organization of rich nodes is crucial because it can
strongly affect our understanding for the whole topological
properties of the network. Our study indicates that in complex
systems the social cohesion (that is the assortativity or
transitivity) of a large community is determined by connectivity
among the leaders (the rich-club). This study also confirms that
although some measures developed in the framework of statistical
physics provide a powerful tool for analyzing the organization of
complex network, in specific situations they are very sensitive to a
small local structure (the connectivity among a very small
rich-club).

Nonetheless, the networks in Table \ref{table1} are not carefully
selected on purpose, and our findings do provide a simple
explanation for the observed properties of many real world networks.
When examining such networks, we need not ask why they exhibit
assortativity or transitivity, but rather how the rich nodes are
connected and why they are connected in this way. For example, in
the case of the Internet the rich nodes form a very strong rich-club
(the various routers are interconnected) and it is this property
that determines the transitivity of the entire network.

Conversely, in some situations (such as to control epidemic spread
or information flow) it is useful to manipulate the assortativity
and transitivity of a large network. Our results provide a cheap and
easy way to do this: just manipulate the connections among the
rich-club members. Followed the work in \cite{Weight_richclub}, an
interesting question to be pursued in future would then be the
investigation of how rich-club affects these important dynamic
processes in weighted and/or directed networks.

This work was supported by the Hong Kong Polytechnic University
Postdoctoral Fellowship Scheme (Fellowships G-YX0N \& G-YX4A). X.-K.
Xu and J. Zhang also acknowledge the National Natural Science
Foundation of China under Grant No. 61004104. The authors thank
Xiang Li and Zhi-Hai Rong for useful discussions and suggestions.


\end{document}